\documentclass[manuscript]{aastex}
\newcommand{\beqa}{\begin{eqnarray}}
\newcommand{\eeqa}{\end{eqnarray}}

\def\hkpc{\ h^{-1}{\rm kpc}}
\def\hmsun{\ h^{-1}M_{\odot}}

\def\kms{{\rm km}\hspace{0.1cm} {\rm s}^{-1}}

\def\vmax{V_{\rm max}}

\def\dhalf{\Delta_{V/2}}
\def\rhalf{R_{V/2}}

\slugcomment{Submitted to ApJ}
\shorttitle{Halo Central Densities}
\shortauthors{Alam, Bullock, \& Weinberg}

\begin{document}

\title{DARK MATTER PROPERTIES AND HALO CENTRAL DENSITIES} 
\author
{S. M. Khairul  Alam\altaffilmark{1} James S. Bullock\altaffilmark{1} 
\& David H. Weinberg\altaffilmark{1}} 
\altaffiltext{1}{Department of Astronomy, The Ohio State University, 
Columbus, OH 43210; alam,james,dhw@astronomy.ohio-state.edu}

\begin{abstract}

Using an analytic model calibrated against numerical simulations, we 
calculate the central densities of dark matter halos
in a ``conventional'' cold dark matter  
model with a cosmological constant (LCDM) and in a ``tilted'' model (TLCDM)
with slightly modified parameters motivated by recent analyses of
Ly$\alpha$ forest data.  We also calculate how  warm dark matter (WDM)
would modify these predicted densities  by delaying halo formation and
imposing phase space constraints. As a measure of central density, we
adopt the quantity $\Delta_{V/2}$, the density within the radius
$R_{V/2}$ at which the halo rotation curve falls to half of its
maximum value, in units of the critical density. 
We compare the theoretical predictions to values of $\Delta_{V/2}$
estimated from the rotation curves of dark matter dominated disk galaxies.
Assuming that dark
halos are described by NFW profiles, our results suggest 
that the conventional LCDM model predicts excessively high dark 
matter densities, unless there is some selection bias in the data toward 
the low-concentration tail of the halo distribution. A WDM model with 
particle mass $0.5-1$keV provides a better match to the observational  
data. However, the modified cold dark matter model, TLCDM, fits the data 
equally well, suggesting that the solution to the ``halo cores'' problem
might lie in moderate changes to cosmological parameters rather than
radical changes to the properties of dark matter.  If CDM halos have
the steeper density profiles found by Moore et al., then neither
conventional LCDM nor TLCDM can reproduce the observed central densities.
\end{abstract}


\section{INTRODUCTION}

A cosmological model based on collisionless  cold dark matter (CDM), a
Harrison-Zel'dovich spectrum  of primordial  density fluctuations, and
cosmological parameters $\Omega_m =  1 - \Omega_\Lambda \approx 0.3  -
0.4$,  $h \approx   0.7$,  can account for   an   impressive range  of
astronomical observations,  especially those that focus on large-scale
predictions  of the theory.   However, on small   scales, there may be
some conflicts.  For  example,   analyses of galaxy rotation    curves
suggest that this model may predict  excessively high densities in the
central regions  of dark matter halos  (Moore  1994; Flores \& Primack
1994; Burkert 1995; Moore et al. 1999;
Navarro  \& Steinmetz 2000; Debattista \& Sellwood
2000; de Blok et al.  2001).

This discrepancy between the theory and observations has  
spurred the exploration of
alternative models  for dark matter.  By stripping
either the collisionless or cold properties  of traditional CDM, or by
considering  additional exotic properties, many  authors have sought to
preserve the  success of   CDM on large  scales  while   modifying the
manifestations of dark matter  on small scales (Spergel  \& Steinhardt
2000; Cen  2000; Peebles 2000; Goodman 2000;  Hogan \& Dalcanton 2000;
Kaplinghat   et al. 2000).   Among   these possibilities,  perhaps the
simplest is  that power is suppressed  on small scales, either because
of a feature in the  initial power spectrum  from inflation 
(Kamionkowski \& Liddle 1999)  or   because of  thermal motion  
of  the  (warm) dark matter particles
(Col\'{\i}n et al.  2000;  Avila-Reese et al.  2000; Eke, Navarro,  \&
Steinmetz 2000; Bode, Ostriker, \& Turok 2000).  A less radical,
and perhaps a more naturally motivated mechanism for suppressing the small 
scale power is to ``tilt'' the initial inflationary
power spectrum to favor large scales.  The effect of such an
adjustment is discussed below.

There are  two   related but  distinct aspects   of the halo   density
problem, the absolute values  of the densities  in the central regions
of halos, and the logarithmic slopes of  the central density profiles.
This paper concentrates on  the first aspect.    As a measure  of 
the 
central density, we advocate the quantity $\dhalf$, the mean density within
the radius $R_{V/2}$ at which the halo rotation curve falls to half of
its maximum  value, in units  of the critical density.  The motivation
for  $\dhalf$   is to focus on   a region that is observationally
robust but in  the  range where conflict with theoretical predictions
may arise.  In  this region
$\dhalf$ can also be probed successfully by density profiles predicted
by N-body simulations.

Using an analytic model calibrated against numerical simulations, we
calculate the central densities in a ``conventional'' cold dark matter
model with a  cosmological constant (LCDM)  and in a tilted LCDM model
with parameters   motivated by recent    analyses of Ly$\alpha$ forest
data (TLCDM).   We also investigate  how warm  dark  matter (WDM) would modify
these predicted densities by delaying  halo formation and by
imposing an upper limit on the central phase space density
(Tremaine \& Gunn 1979).
For  each of
these  models  we  compare the  predicted values   of $\dhalf$ to  the
observationally inferred  central  densities obtained from  a compiled
list of dark-matter dominated dwarf and LSB galaxies.

The rest of the paper is organized as follows.  In the next section we
describe our density  parameter, $\dhalf$, and present observationally
inferred values   of $\dhalf$ for our  list  of  dark-matter dominated
galaxies.  In  \S 3,  we  present the predictions  of our adopted
cosmological models for  halo densities. The  theoretical calculations
for the central densities in the conventional LCDM model
and the tilted TLCDM model are presented in
\S 3.1. The  warm dark matter  (WDM) predictions based on delayed
halo formation and phase space constraints are discussed in \S 3.3 and
\S 3.4, respectively.  We discuss the comparison between the 
theoretical predictions and the observational results in \S 4.

\section{CHARACTERIZING HALO CORE DENSITIES}

The density profiles of dissipationless dark matter halos formed in
 high-resolution N-body simulations generally show
central density cusps, $\rho \propto r^{- \gamma}$.  
Whether the asymptotic central slope is
$\gamma \simeq 1$, as suggested by the analytic form of 
Navarro, Frenk, \& White (1996, hereafter NFW),
\begin{equation}
\rho_{\rm NFW}(r) = \frac{\rho_0}{(r/r_s)[1+r/r_s]^2},
\end{equation}
or $\gamma \simeq 1.5$ as suggested by Moore et al. (1999),
\begin{equation}
\rho_{M}(r) = \frac{\rho_0}{(r/r_M)^{1.5}[1+r/r_M]^{1.5}},
\end{equation}
is still subject to some debate (see, e.g., Klypin et al. 2000).  
Since CDM-derived profiles appeared to predict too much  
mass at small scales, 
Burkert    (1995) proposed an
alternative density  profile with a  flat  density core to  model  the
observed rotation curves of several dark matter dominated galaxies,
\begin{equation}
\rho_{B}(r)=\frac{\rho_{0}}{(r/r_{B}+1)[r^2/r_{B}^2+1]}.
\end{equation}  
For comparison,  we  will also consider the popular 
truncated isothermal
profile with a core,\footnote{Shapiro, Iliev \& Raga (1999) have a simple
model for the postcollapse
equilibrium structure of CDM halos that predicts halo profiles of
approximately this form.}
given by
\begin{equation}
\rho_{{\rm Iso}}(r)= \frac{\rho_{0}}{(1+ r/r_c)^{2}}.
\end{equation}   
For brevity, we will refer to equations (1)-(4) as NFW, Moore,
Burkert, and Iso+core profiles, respectively.

One of the challenges   in following the  controversial topic  of
halo densities is that
theoretical profiles and observed rotation curves have been fit with a
variety of different analytic models.  The  degree of conflict between
theory  and  observations  often depends  on   the  way in  which  the
comparison is made.  As  a simple but  (almost) observable measure  of
halo core densities, we propose the quantity
\begin{equation}
\dhalf = {\bar{\rho}(R_{V/2})\over \rho_{\rm crit}} = 
               {1\over 2}\left({V_{\rm max}\over H_0 \rhalf}\right)^2  = 
               5\times 10^5 \left({V_{\rm max}\over 100 \;\kms} \right)^2
               \left({1\hkpc\over \rhalf}\right)^2,
\end{equation}
the ratio of the mean dark matter density to the critical density 
$\rho_{\rm crit}=3H_0^2/8\pi G$ within the radius $R_{V/2}$ at which the 
galaxy rotation curve falls to half of its maximum value $V_{\rm max}$. 
One can also interpret  $\dhalf$ in terms of the number of rotation
periods per Hubble time at the radius  $\rhalf$:
\begin{equation}
\ N_{\rm rot} = {V_{\rm max}/2\over 2 \pi \rhalf H_0 }  = 
                {\left(\dhalf\over 8\pi^{2}\right)^{1/2}.}
\end{equation}

Our focus on the  mass density within a fiducial  radius is similar to
that of Navarro \&   Steinmetz  (2000). 
However, by using the    radius
``halfway up'' the rising part of the rotation curve, we concentrate on
the  region  where   conflicts  between  predicted and  observed  halo
densities are  more severe (and  where the  observations are still typically
robust  to  resolution   uncertainties).     Accurate measurement   of
$\Delta_{V/2}$ requires subtraction  of the baryon contribution to the
rotation  curve, which is  most feasible  in the  case  of low surface
brightness  galaxies.   

Figure  1   shows rotation  curves   for  different  density  profiles
normalized  to the same $V_{\rm  max}$  and $R_{V/2}$.   Though we will be
focusing  on  values of $\dhalf$  ,  different  models do have  different
shapes at  fixed $V_{\rm  max}$ and  $R_{V/2}$.  The shapes   could be
distinguished,  for  example, by measuring   the logarithmic slopes at
$\rhalf$.  For the four curves shown, the log-slopes are approximately
0.76, 0.62,  0.45,  and 0.25 for   Burkert, ISO+core, NFW, and Moore 
profiles, respectively.  These slopes would be distinguishable with
sufficiently good data, but slope measurements place much stricter demands
on the accuracy of corrections for baryon contributions and for
non-circular motions.

Points in Figures 2 and 3 show  values of $\dhalf$ and $\rhalf$ versus
$V_{\rm max}$ derived   from  the observational analyses of    Burkert
(1995), Kravtsov et al.  (1998), and Swaters  et al. (2000).  
Points in Figures 4 and 5 show the corresponding values of
$R_{V/2}$, which range from $\sim 0.5\hkpc$ to $\sim 5\hkpc$.
In their analyses, Burkert (1995) used equation (3) to fit 
dwarf galaxy rotation curves,
Kravtsov et al. (1998) used a different density profile 
to fit dark matter dominated dwarf and LSB galaxies, and 
Swaters  et al. (2000) used the  Iso+core profile
to model dwarf and  LSB galaxies.~\footnote{Some of  the galaxies in these
samples overlap; however, we represent the  derived $\dhalf$ values by
separate points for separate  analyses.}     For the Burkert 
galaxies, we use rotational velocities   and core radii
given in his figure 2.  For the Kravtsov et al.  (1998) sample, we use
their advocated  velocity profile  (see their  equation 4) 
with parameters provided in their table 2.
For the
Swaters  et al.   (2000)  data,  we use  the core  radii  and central
densities from their table 2 for the minimum disk mass model. 
One of the advantages of focusing on $\dhalf$ is that it
can be estimated from the parameters of any model that fits the
rotation curve data, allowing us to combine observational results
from several sources without refitting the original measurements.
Indeed, one can estimate $\dhalf$ by visual inspection from a 
plotted rotation curve that extends far enough to allow determination
of $\vmax$.  We have examined the high resolution optical rotation
curves of McGaugh et al.\ (2001), and for the seven galaxies
that permit accurate determination of $\vmax$ and $\rhalf$ we
find six values of $\dhalf$ in the range $10^5-10^6$,
in agreement with the data points plotted
in Figures 2 and 3.  (One galaxy has $\dhalf\simeq 5\times 10^6$.)

There are many potential  complications associated with a  theoretical
comparison to rotation curve data.   One is  that the predictions  for
shapes  of   dark halo    density   profiles (\S     2) are  based  on
dissipationless simulations, and therefore do not include the effect that
baryonic infall   and disk formation will  have  on the final rotation
curve of the  galaxy.  In  order  to minimize this  problem,  
we  have restricted ourselves   to galaxies that appear to be
dark matter dominated.   Among the complete  lists of  rotation curves
presented in the samples  described above, we use  those for which the
estimated dark matter  contribution to the  rotation curve at $\rhalf$
is greater than $80\%$.

An additional uncertainty  is  associated  with the  determination  of
$V_{\rm max}$, which, depending  on  the underlying  density  profile,
requires measurement  of  the rotation curve   to fairly large radius.
For example, when matched to coincide at small radii, as in Figure 1,
the Moore profile rotation curve continues to rise
out to larger radii compared to the others, and therefore the value of
$V_{\rm max}$ is likely to be  systematically underestimated.  This is
not as much of a  problem if the underlying profile  is of the Burkert
form, since the rotation  curve generally reaches a  maximum at a much
smaller radius.  This effect is systematic: the more centrally
peaked the underlying density profile, the more likely it is 
that the value of $V_{\rm max}$ will be underestimated.

Errors in determining $V_{\rm max}$ translate into 
uncertainties in the values of $\rhalf$ and $\dhalf$.  
If the   rotation  curve   is  linear  (solid   body)  near
$R_{V/2}$, errors  in the determination  of   $V_{\rm  max}$ and
$R_{V/2}$    have    canceling  effects  in    the    determination of
$\Delta_{V/2}$ (see eq. 5).
As the log-slope of the rotation curve flattens,
the implied uncertainties in $\dhalf$ increase.
For example, if $V_{\rm max}$ is underestimated by
10\%, then the corresponding value of $\dhalf$
will be {\it over}estimated by factors of 1.02, 1.12, 1.44, and 2.11
for the Burkert, Iso+core, NFW, and Moore profiles,
respectively.  The arrows in Figure 2 illustrate how the
data points  should be shifted if
$V_{\rm max}$ has been underestimated by 10\%, under the assumption
that the underlying profiles are (left to right) NFW,
Moore, Burkert, and Iso+core.

\section{PREDICTIONS OF COSMOLOGICAL MODELS}

One popular way to characterize  halo density profiles  is in terms of
their concentrations, i.e., 
$c \equiv R_{\rm  vir}/r_s$ for  NFW,
$c_{M}\equiv  R_{\rm vir}/r_M$
for Moore, 
$c_{c} \equiv R_{\rm vir}/r_c$ 
for  ISO+core,  
and  $c_B \equiv R_{\rm vir}/r_B$ for Burkert.
Here $R_{\rm vir}$ is  the virial radius defined  as the radius within
which the mean density is 337 (for our assumed $\Lambda$ cosmology) 
times the average density of  the    Universe. For  the class    of cosmologies
considered here, $R_{\rm vir}$ is related  to the halo virial mass and
virial velocity at $z=0$ via
\begin{eqnarray} 
M_{\rm vir} & \simeq & 5 \times 10^{10} \hmsun
\left(\frac{R_{\rm vir}}{75 \hkpc}\right)^3,  
\end{eqnarray}
and
\begin{eqnarray} 
V_{\rm vir} & \simeq & 60 \; \kms  \left(\frac{R_{\rm vir}}{75 \hkpc}\right).
\end{eqnarray}
The concentration parameters are defined in  terms of specific density
profiles,  but  they can be related  to  the general parameters
$\dhalf$, $\rhalf$, and $V_{\rm  max}$.  For NFW  these parameters
are
\begin{equation}
\rhalf = 0.13 r_s,~ \dhalf = 409 c^3/f(c),~ V_{\rm max}^2 
\simeq 0.2 V_{\rm vir}^2 c/f(c),
\end{equation}
where $f(c) = \ln(1 + c) - c/(1+c)$. Similar  transformations
allow us to normalize the implied rotation 
curves of our other density profiles.
For the Moore profile we have
\begin{equation}
\rhalf = 0.03 r_M,~ \dhalf = 1.56 \times 10^{4} c_M^3/f(c_M), 
~V_{\rm max}^2 \simeq 0.47 V_{\rm vir}^2 c_M/f_M(c_M),
\end{equation}
where $f_M(c_M) = \ln(1 + c_M) - c_M/(1+c_M)$.  For the 
Iso+core case we have
\begin{equation}
\rhalf = 1.13 r_c,~\dhalf = 25 c_c^3/{(c_c-f_c(c_c))}~,
~V_{\rm max}^2 \simeq V_{\rm vir}^2c_c/{(c_c-f_c(c_c))},
\end{equation}
where $f_c(c_c) = \tan^{-1}c_c$. For the Burkert profile
\begin{equation}
\rhalf = 0.5 r_B,~\dhalf = 119 c_B^3/f_B(c_B),~
V_{\rm max}^2 \simeq 0.86 V_{\rm vir}^2c_B/f_B(c_B),
\end{equation}
with $f_B(c_B) = \ln((1 + c_B)^2(1+c_B^2)) -2\tan^{-1}c_B$.

\subsection{LCDM models}

Analyses such  as  those presented in   NFW and Bullock  et al. (2001;
hereafter B01) suggest that the central  densities of halos are set by
the density of the universe at the characteristic collapse time of the
halo.   B01 proposed  an analytic  model   in which the halo  collapse
redshift, $z_c$, is given by the epoch at which  the mass scale of its
subunits was first going  non-linear and  is
therefore  directly linked to the amplitude   of the power spectrum on
small scales (see B01 for details and Wechsler et al.\ [2001] for
an improved model along similar lines).
In  this model, a halo observed  at redshift $z_0$  is
expected to have  an NFW concentration $c  \simeq 4 (1  + z_c)/(1 +  z_0)$.
B01 showed  that  this simple  prescription  accurately reproduced the
median $c$  values of halos  as  a function  of  mass and redshift for
models with cold dark matter and power-law initial  power spectra.  In
the ``standard''     LCDM  model   with  parameters     $\Omega_m  = 0.3$,
$\Omega_\Lambda =  0.7$, $h=0.7$, and   $\sigma_8 = 1.0$, we  use  the
result of B01  for $c$ as a function of $M_{\rm vir}$, then  
translate to the predicted parameters 
$\dhalf$ and $\rhalf$ for NFW profiles as discussed in the previous section.
The trends of these
parameters  are shown in Figures  2  and 3. The  error  bar in Figure 2
shows the $\pm  1\sigma$  scatter in concentrations  (translated  to a
scatter in $\dhalf$) found in N-body simulations by B01 (see also
Wechsler et al. 2001).  We will discuss the comparison to observations
in \S 4 below, but it is immediately obvious that the mean LCDM
prediction is well above the median of the data points.

Since CDM halo densities are closely linked to the power
spectrum amplitude, it is interesting to consider how simply changing
parameters within the CDM picture will modify predictions.
As an example, we take the model favored by recent
 Ly$\alpha$ forest observations (McDonald et al. 2000; 
Croft et al. 2001): $\Omega_m = 1 - \Omega_\Lambda = 0.4$,
$h=0.65$, $\sigma_8 = 0.66$, and an inflationary ``tilt'' of
$n=0.9$.   In this TLCDM model, the power
is significantly reduced on galactic and sub-galactic scales 
relative to the LCDM model.  This leads to substantially 
lower central densities, about a factor of four in $\dhalf$,
as shown by the thin black lines
in Figures 2 and 4.  These analytic predictions agree with
N-body simulations of the TLCDM model (Kravtsov, private communication).

As discussed in \S 2, whether CDM halos more closely resemble
a Moore profile or an NFW profile remains an open question.
Although models like that of B01  provide a good
characterization of general halo density structure on the
scale of $r_s$, the implied values of $\dhalf$ depend
on how the profiles are extrapolated to small scales.  In order
to illustrate this point, and to allow for the possibility
that CDM halos more closely follow the Moore profile,
the dashed lines in Figure 2 show how the predictions change
for the Moore profile assumption.  We have calculated the
change using the results of Klypin et al. (2000), who found that
when the same halo is fit using NFW and Moore profiles,
the best-fit $\vmax$ values are nearly identical and the 
scale radii obtained are related as $r_M \simeq 2 r_s$.  
Using this relation, the implied shift in our central
density parameter is $\dhalf^{M} \simeq 4 \dhalf^{\rm NFW}$;
a Moore profile extrapolation predicts central densities that
are four times higher.

\subsection{WDM models: delayed halo formation}

One proposed solution    to   the central density   crisis    involves
substituting  warm dark matter (WDM)  for  CDM.  In principle, WDM can
help lower halo densities for two reasons.  First, phase space
considerations for fermionic warm dark matter impose upper limits
on the WDM central densities, as discussed in
in the next section.  The second  effect is a result of WDM
free streaming, which washes out    fluctuations on small scales   and
delays the collapse of the low-mass  subunits that make up galaxy-size
halos.   As  discussed above,  later collapse  is expected  to imply
lower central densities.

Using simple models like those discussed in the previous section,
one would expect WDM halo concentrations to be lower
than  those of similar mass LCDM halos, as a result
of the suppressed  power spectrum amplitude on small scales.
Several authors have used     N-body   simulations to verify      this
expectation  (Col\'{\i}n et  al. 2000; Bode,   Ostriker, \& Turok 2000;
Eke, Navarro  \& Steinmetz 2001;  hereafter  ENS).  ENS
found that the central  densities  in their WDM  halos were
suppressed  even   relative to the   B01  model predictions,  and this
motivated them to propose a modified analytic model  in which the halo
formation time is based on the  magnitude {\it and} slope of the
power   spectrum (see ENS  for   details).  The  long-dashed lines  in
Figures 3 and  5 show the ENS model  predictions for three WDM models.
In each  case, we have  assumed  the same parameters  as  those in our
standard  LCDM case with the added  assumption that the dark matter is
in the form  of a light  fermion with mass  $m = 0.2\;$keV (thin lines),
$0.5\;$keV  (medium lines), and  $1\;$keV  (thick lines).
Unfortunately,  the  ENS analytic  model  has  not been tested
against N-body simulations over the range that seems most relevant for
the data plotted in Figures 3 and 5, and it is for these lower 
$V_{\rm max}$ values that it predicts substantial decreases in halo
central  densities.  For this reason,  we  also include the B01  model
predictions  in Figures 3 and   5, displaying them with solid lines.
One   may conservatively  expect the properties of  WDM halos 
with a given particle mass to lie between the corresponding
long-dashed and solid lines.

The predictions in Figures 3 and 5 assume that WDM models produce
halos with NFW profiles, and that the effect of suppressed small scale
power is simply to lower the concentration parameters as predicted
by the ENS or B01 models.  This assumption is consistent with 
existing numerical results, but it is not clearly mandated by them.
As for the LCDM models, our predicted densities would rise by 
a factor of four if we assumed Moore profiles rather than NFW profiles.

\subsection{WDM models: phase space constraints}

WDM  also   imposes  a maximum central  phase   space  density by  the
argument first advanced by Tremaine \& Gunn (1979).
The basis  of  the  Tremaine-Gunn
argument is  that the maximum coarse-grained phase space  density of halo dark
matter   cannot exceed  the maximum  primordial   
fine-grained phase space density (which is conserved for collisionless
matter).
A primordial neutrino gas follows  a Fermi-Dirac distribution, 
with an occupation number $f$  that reaches a maximum
of $f_{\rm max}=0.5$, implying a maximum phase space density
of $g_{\nu}/h^{3}$.
In order to constrain the central density, we make the
assumption that the mean distribution of the WDM fermions in the
final collapsed halo has a Maxwellian form~\footnote{This is
a reasonably good approximation for CDM dark matter halos formed
in N-body simulations (Vitvitska et al. 2001)}
\begin{equation}
\langle f(p)\rangle=f_{\max} \exp(-v^2/{2\sigma^{2})}.
\end{equation}
In this case, the central density of the particles in the dark halo 
is
\begin{eqnarray}
\rho_0  = 2\frac {g_{\nu}}{h^{3}} m_{\nu}\int\langle f(p)\rangle d^{3}p 
= \frac{g_{\nu}}{h^{3}}m_{\nu}^4(2\pi\sigma^2)^{3/2},
\end{eqnarray}
where $m_{\nu}$ is the mass of the fermionic particle
and $\sigma$ is the one dimensional velocity dispersion.
%The  factor  $(g_{\nu}/{h^{3}})m_{\nu}^4$ in  the central  density  is
%simply the  maximum phase  space density  in  $m^4_{\nu}$ units,
%which depends  only on the  particle properties like the particles are
%thermal or degenerate, bosons  or fermions. 
The maximum phase  space mass-density, 
$Q \equiv \rho_0/(2 \pi \sigma^2)^{3/2}$,
for $g_{\nu}=2$ is then\footnote{ This phase
space density is  higher by a  factor of 2 than  the value given by
Hogan and Dalcanton (2000) because    we use  both the particle    and
anti-particle properties of fermions.}  
\begin{equation}
\ Q_{\rm max}\simeq
   10^{-3} 
   \frac {M_{\odot}/\rm pc^{3}}{(\kms)^3}(m_{\nu}/1\rm keV)^{4}.
\end{equation}

Recently,  Hogan  and  Dalcanton (2000) used this Tremaine-Gunn
limit on $Q$ to determine core
radii for dark matter  dominated galaxies by  assuming Iso+core
halo density   profiles. We  instead use   
the Burkert profile (eq. 3),
which we expect to provide a more realistic model for a WDM
halo.  On scales where the primordial phase space constraint is
unimportant, WDM should behave much like CDM, and outside of 
the constant density core the Burkert profile resembles the 
NFW form found for CDM.

The complication of using a Burkert profile for the halo is its
non-isothermal velocity dispersion --- to impose the maximum phase
space density constraint, we must use the central velocity dispersion
rather than the virial velocity dispersion.
For a spherical system, the Jeans equation reduces to
\begin{equation} 
\ \frac{1}{\rho(r)}\frac{d(\rho\sigma^{2})}{dr}= - \frac{d\Phi}{dr}.
\end{equation}
By setting ${d\Phi}/dr= GM(r)/r^{2}$ and assuming a Burkert $M(r)$ profile,
one obtains, after some manipulation, an attractively simple formula
for the central velocity dispersion:
\begin{equation}
\ \sigma_0^2 = K\, V_
{\rm max}^2\int_{0}^{\infty}\frac{f(x)}{x^{2}(1+x)
(1+x^{2})}dx \simeq 0.2 V_{\rm max}^2,
\end{equation}
where the constant $K \simeq 1.2$.
This allows us to write the central phase space density as
\begin{equation}
 Q_{B} = \frac{\rho_B}{(2 \pi \sigma_0)^{3/2}} 
\simeq 2.7 \times10^{-8} \left( \frac{V_{\rm max}}{10\; \kms} \right)^{-3}
\left(\frac{h}{0.7}\right)^2 
\frac{c_B^3}{f_B(c_B)} 
\frac {M_{\odot}/{\rm pc}^3}{(\kms)^3}.
\end{equation}
For a given $V_{\rm max}$ and WDM particle mass $m_{\nu}$, we obtain the 
Burkert concentration $c_B$ by  equating $Q_B$ 
to  the maximum allowed value $Q_{\rm max}$. 
We then compute $R_{V/2}$ and $\Delta_{V/2}$ from the
density profile (eq. 12).  Results
are shown by the short-dashed lines in Figures 3 and 5 and
discussed below.

\section{DISCUSSION}

The observational data plotted in Figures 2 and 3 show typical values
$\Delta_{V/2}$   in the range $10^5-10^6$, or a typical ``rotation rate''
of $\sim 80$ rotations per Hubble time at $\rhalf$ 
(eq. 6 for $\dhalf=5\times 10^5$).
There  is no clear trend
between $\Delta_{V/2}$ and $V_{\rm max}$, but there is substantial scatter.
Some  of this scatter could of course be observational in origin,
and some could arise from the residual influence of baryons.
However, the scatter  is comparable to the physical scatter
found   for simulated  halos by  B01  (error bar  in   Figure 2), 
which arises from    the  variation  in halo formation
histories. 

Any comparison between theoretically predicted halo densities
and these observational data rests on a critical implicit assumption:
the baryons have not significantly altered the dark matter profile
itself.  Since the most obvious impact of baryon dissipation is
to draw the dark matter inward (Blumenthal et al.\ 1986), it is
generally believed that baryons will only increase halo central
densities.  However, more complicated baryon/dark matter interactions
could have the opposite effect, for example by compressing dark 
matter adiabatically and releasing it impulsively after a galactic
wind (Navarro, Eke, \& Frenk 1996), or by altering dark matter
orbits through resonant dynamical interactions with a rotating bar
(N.\ Katz \& M.\ Weinberg, private communication; see
Hernquist \& Weinberg 1992).  Because the galaxies in our
sample are chosen to be dark matter dominated (with at least
80\% of the circular velocity at $\rhalf$ contributed by the
dark halo), we will assume that baryonic effects on the
dark matter profile have been negligible in these systems.
This assumption appears reasonably safe, but not incontrovertible.

The  LCDM model predicts  a mean $\Delta_{V/2} > 10^6$ for
$V_{\rm max} \la 100\;\kms$.  It is ruled out by the observational data
{\it  unless} the LSB  galaxies  selected for analysis preferentially
sample the low density tail of the halo population.  If the halo
profiles are better represented by the Moore form, then
the disagreement becomes even worse.  We therefore concur with
previous arguments that the LCDM model with these parameter choices
predicts excessively high dark matter densities in the central
parts of halos.

A WDM particle mass of $0.5-1.0$  keV can yield  good agreement with the
median observed  $\Delta_{V/2}$, as shown in Figure 3.
Unfortunately, the analytic  models
have not been tested numerically in 
the observationally relevant range of $V_{\rm max}$,  
so it is  hard to say just  what particle mass is required 
to match the data --- 0.5 keV seems best if we adopt the B01 model
for halo concentrations, 1 keV if we adopt the ENS model.
In the $\vmax$
range probed  by  current data, the  maximum  values of $\dhalf$ allowed by
the phase space argument of \S 3.3
are higher  than those predicted by formation
time  arguments of \S 3.2.  This result has two related implications.
First, on scales $\sim \rhalf$, the densities of WDM halos will
be determined by formation time rather than by phase space constraints,
except for halos with very low $\vmax$.  Second, the radii of the
constant density cores in WDM halos will be much smaller than $\rhalf$,
and they will typically be too small to have a measurable impact on
rotation curves.  Therefore, WDM models are likely to be no more (or less)
successful at matching observed rotation curves than CDM models
that have similar NFW halo concentrations.

This last remark leads us back to Figure 2 and to our most intriguing
quantitative result: the TLCDM model predicts mean halo densities near the
median of the observational values, similar to those of 0.5--1 keV WDM halos,
and about  a factor of  four lower than  those of the conventional  LCDM model.
This result suggests that the reconciliation of predicted and observed
halo densities may require only moderate adjustments to the parameters
of the  inflationary cold dark matter model,  not a  radical change in
the properties of dark matter particles.  If the true halo profiles have
the Moore form rather than the NFW form, then WDM and TLCDM both
have trouble reproducing the data, though they still come substantially
closer than LCDM.

Among our two  successful models, the TLCDM model  is much more in line
with the standard scenario of structure formation, and it does not
require tuning of a model parameter (the particle mass) specifically
to match the observed halo densities.
The TLCDM model explored here has cosmological
parameter values that  are favored by  a combination of COBE microwave
background  measurements and the  matter  power spectrum inferred from
Ly$\alpha$ forest data  (see Croft et  al.\ 2001).  The assumed degree
of tilt (we used $n=0.9$) is not unrealistic, but  a natural
outcome of both small-field and  large-field inflation models, as well
as a  subset of ``hybrid'' inflation models   (see, e.g., Hannestad et
al.  2001).  A primordial slope  of this type  is consistent with  the
latest CMB  measurements, and it is favored   by the CMB  plus large scale
structure analysis of Wang, Tegmark, \& Zaldarriaga (2001), who obtain
$n \simeq 0.91 \pm 0.1$.  Although our adopted TLCDM model is somewhat
disfavored by the observed cluster  abundance,  producing a number  density  
of clusters  roughly $2  \sigma$ below  the  estimated value  from Eke et
al. (1996), the ``standard'' LCDM model is similarly disfavored by the
Ly$\alpha$  forest.  Since  standard  LCDM seems to   be facing severe
problems  on small   scales,  TLCDM may represent    a natural way  to
reconcile theory with observation,  without relaxing any (more) of the
desirable aspects of the CDM paradigm.

\acknowledgments

We thank Andrey Kravtsov for helpful discussions of halo concentrations
and profiles, Douglas Richstone for pointing out the connection between
$\Delta_{V/2}$ and rotation rate, and Robert Scherrer for helpful input
during the early phases of this project.   This work was supported
in part by NSF grant AST-9802568.

\clearpage

{\begin{figure}[t]
\vspace{2.5in}
\label{FIGURE1}
\includegraphics{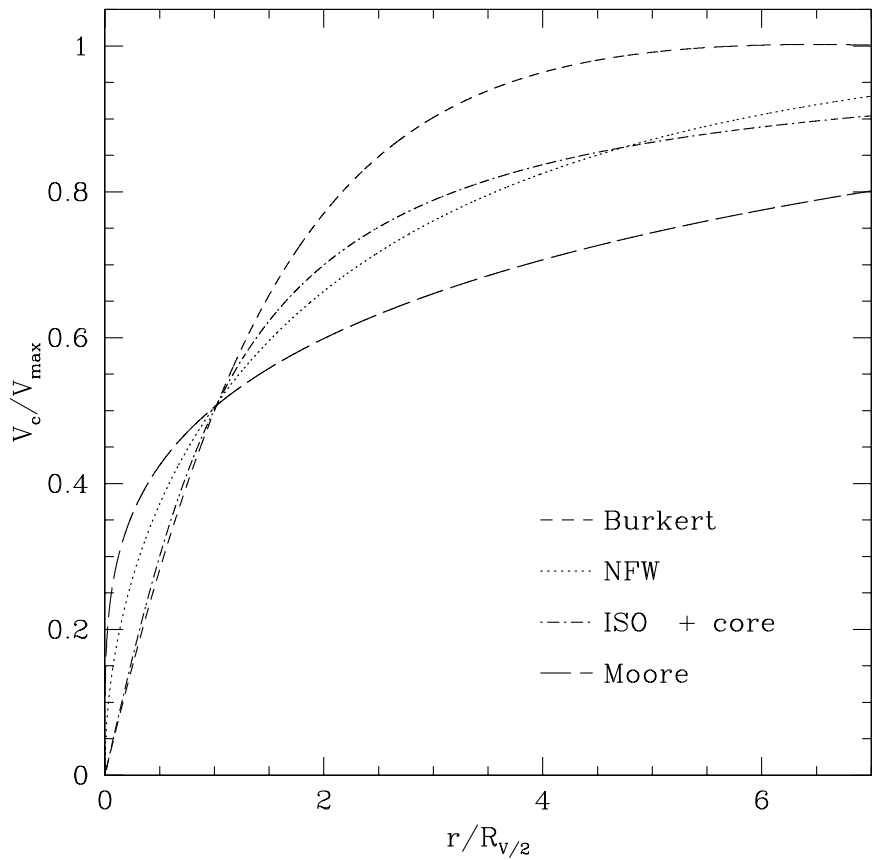}
\figcaption{Rotation  curves   predicted by the NFW, Moore, Burkert,
and Iso+core density profiles (eqs. 1-4, respectively).
Velocities are scaled to the maximum circular velocity of the
halo, $\vmax$, and radii are scaled to the radius $\rhalf$ 
at which $V_c=0.5\vmax$.  In this scaled form, the rotation curves
are independent of the halo concentration parameters.
\label{FIGURE1}}
\end{figure}}

\clearpage

{\begin{figure}[t]
\vspace{2.5in}
\label{FIGURE2}
\includegraphics{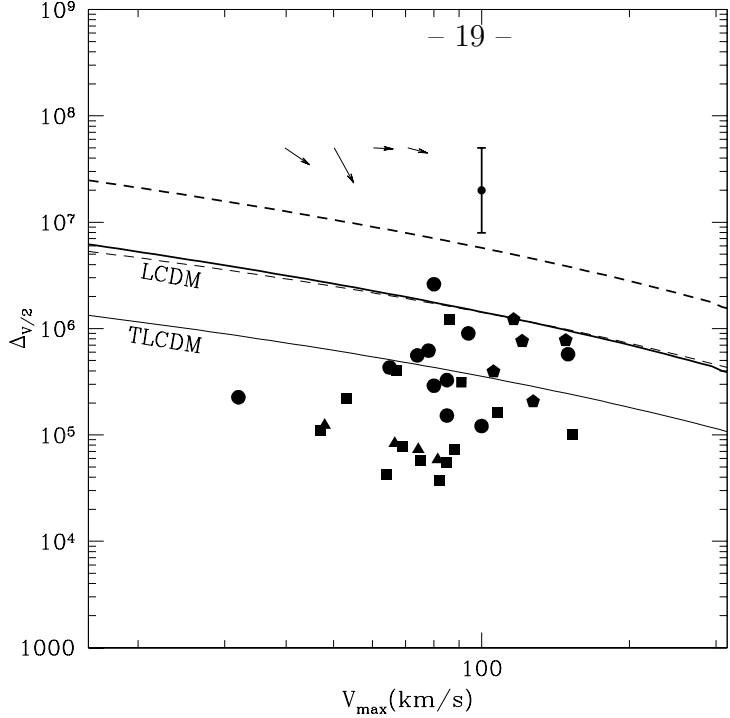}
\figcaption{
Halo central densities as a function of maximum circular velocity,
where $\dhalf$ is the ratio of the mean density within $\rhalf$ to 
the critical density (see eq. 5).  Points show values derived 
from the rotation curves of dark matter dominated galaxies, using
the samples and fits of Burkert (1995) (triangles), Kravtsov
et al. (1998) (squares), and Swaters et al. (2000) (circles).  
Heavy and light solid lines show $\dhalf$ predictions for the 
LCDM and TLCDM models, respectively, obtained by combining the 
B01 model for halo concentrations with an assumed NFW profile.
Dashed lines show corresponding results assuming a Moore profile.
The error bar shows the 1-$\sigma$ halo-to-halo scatter in $\dhalf$
predicted by the B01 model.  Arrows illustrate the impact of a 10\%
$\vmax$ error on the observed data points in the case where the underlying
halo has a (left to right) NFW, Moore, Burkert, or Iso+core profile;
if $\vmax$ for a given data point is underestimated by 10\%, then it
should be shifted in the direction and amount indicated by the arrow.
\label{FIGURE2}}
\end{figure}}

\clearpage

{\begin{figure}[t]
\vspace{2.5in}
\label{ FIGURE 3}
\includegraphics{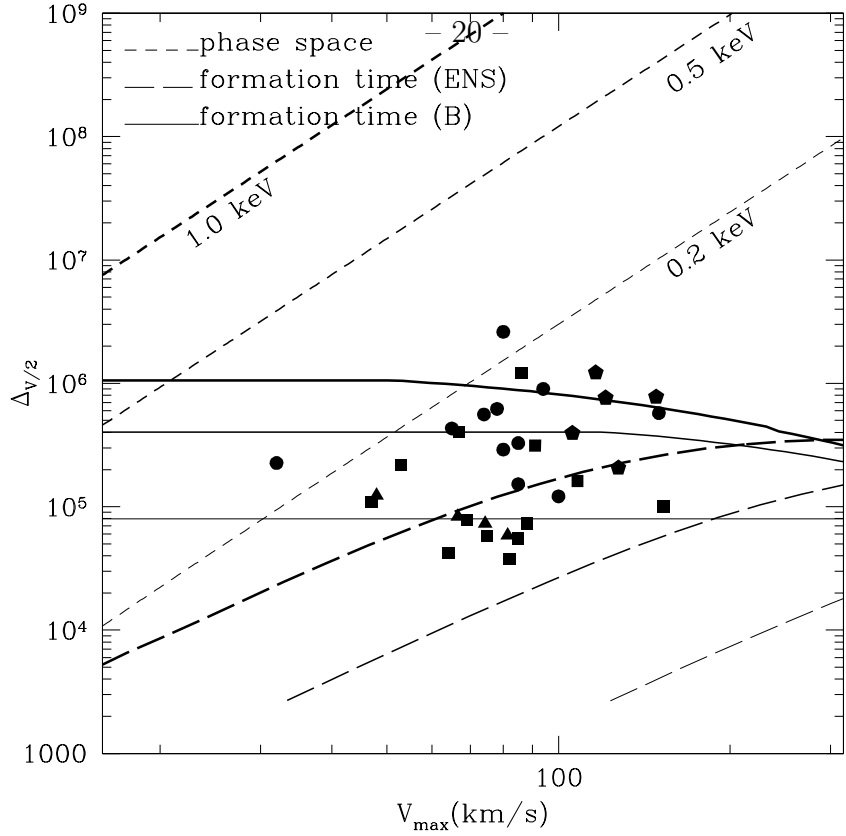}
\figcaption{
Halo central densities in the WDM model.
Points are the same as those in Figure 2.
Solid lines show values of $\dhalf$ predicted by the B01
formation time calculation for a WDM particle mass of
1 keV (heavy), 0.5 keV (medium), or 0.2 keV (light).
Long-dashed lines show predictions for the same models
using the ENS formulation, which yields lower halo concentrations.
Short-dashed lines show the maximum values of $\dhalf$ allowed
by phase space constraints, as described in \S 3.3.
These lie well above the values predicted by formation time
arguments except when $\vmax$ is very low.
\label {FIGURE 3}}
\end{figure}}

\clearpage

{\begin{figure}[t]
\vspace{2.5in}
\label{ FIGURE 4}
\includegraphics{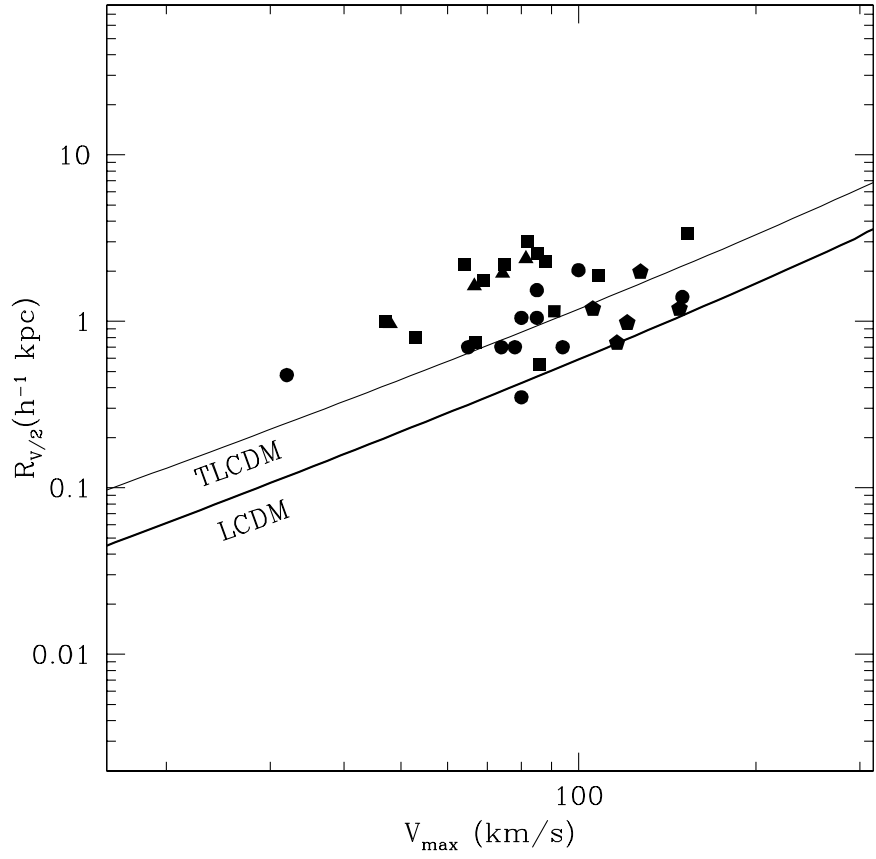}
\figcaption{  
Relation between $\rhalf$ and $\vmax$ in the LCDM and TLCDM models,
computed in the same way as the $\dhalf$ predictions of Figure 2.
Points show the same observational data as Figures 2 and 3,
now translated to the $\rhalf$---$\vmax$ plane.
\label {FIGURE 4}}
\end{figure}}

\clearpage

{\begin{figure}[t]
\vspace{2.5in}
\label{ FIGURE 5}
\includegraphics{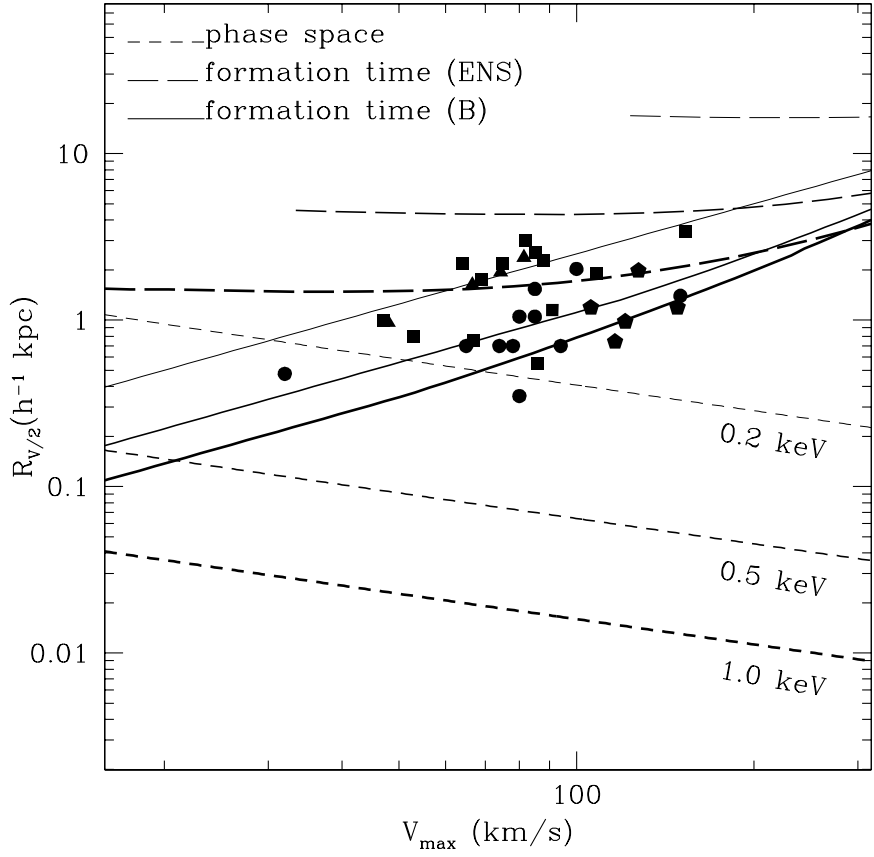}
\figcaption{
Like Figure 4, but for WDM models, with predictions calculated as
described in the Figure 3 caption.
\label {FIGURE 5}}
\end{figure}}

\end{document}